\documentclass[12pt,single]{article} 
\usepackage{anysize}

\usepackage[dvips]{graphicx} 
\usepackage{amsmath}
\usepackage{amssymb} 
\usepackage{amsfonts}
\usepackage{dcolumn}
\usepackage[hang,small,bf]{caption}

\marginsize{2.5cm}{2.5cm}{2.1cm}{2.5cm}  
\begin{document}

\begin{center}
\section*{Lamb   Shift in Muonic Hydrogen  }
\end{center}    

\subsection*{E. Borie}

\subsubsection*{Forschungszentrum Karlsruhe,      \\                     
Institut f\"ur Hochleistungsimpuls and Mikrowellentechnik (IHM), \\ 
Hermann-von-Helmholtzplatz~1,\\ 76344 Eggenstein-Leopoldshafen, Germany}

\bigskip
PACS Numbers  36.10.-k;  36.10.Dr;  31.30.Jv. 

\vspace{1.2cm}

\subsubsection*{Abstract}
\vspace{-0.2cm}
The Lamb shift in muonic hydrogen continues to be a subject of
experimental and theoretical 
investigation.  Here my older work on the subject is updated 
to provide a complementary calculation of the energies of the 2p-2s
transitions in muonic hydrogen.


\subsubsection* { Introduction }  
\vspace{-0.2cm}
The energy levels of muonic atoms are very sensitive to effects of
quantum electrodynamics (QED), nuclear structure, and recoil, since the
muon is about 206 times heavier than the electron \cite{RMP}.  In view
of a proposed measurement of the Lamb shift im muonic hydrogen
\cite{experiment}, an improved theoretical analysis seems to be
desirable.  Since the first theoretical analysis \cite{digiacomo}, the
subject of the Lamb shift (the 2p-2s transition) in light muonic atoms
has been investigated with increasing precision by a number of  authors 
\cite{Borie75,Borie78,hyperfine,BorieHe3,Pachuki1,Pachuki2,eides}.  
The present paper provides an independent recalculation of some of the
most important effects, including hyperfine structure, and a new
calculation of  some terms that were
omitted in the most recent literature, such as the virtual Delbr\"uck
effect \cite{Borie76}.   An alternative calculation of the relativistic
recoil correction is presented. 

In the numerical calculations the fundamental constants from the CODATA
 1998 (\cite{Mohr}) are used, i.e.: 
$\alpha^{-1}$, $\hbar c$, $m_{\mu}$, $m_e$, $m_u$\,=\,137.0359998,
 197.32696\,MeV$\cdot$fm, 
 105.658357\,MeV, 0.5109989\,MeV,  931.4940\,MeV, respectively.     
The changes in these constants in the CODATA 2002 compared with CODATA
1998 are too small to make any relevant difference in the results.

\subsubsection*{Vacuum Polarization }
\vspace{-0.2cm}
The most important QED effect for muonic atoms is the virtual production
and annihilation of a single $e^+ e^-$ pair 
It has as a consequence an effective interaction of 
order $\alpha Z \alpha$ which
is usually called the Uehling potential (\cite{Uehling,serber}.  This
interaction describes the most important modification of Coulomb's law.
Numerically it is so important that it should not be treated using
perturbation theory;  instead the Uehling potential should be added to
the nuclear electrostatic potential before solving the Dirac equation.
However, a perturbative treatment is also useful in the case of very
light atoms, such as hydrogen.  

Unlike some other authors, we prefer to  use relativistic (Dirac) wave
functions to describe  the muonic orbit.  This is more exact, and as
will be seen below, it makes a difference for  at least  the most
important contributions.  The wave functions are given in the book of
Akhiezer and Berestetskii \cite{akhiezer} and will not be given here.       
In perturbation theory, the energy shift due to an effective potential 
$\Delta V$ is given by 
\begin{equation}
 \Delta E_{n \kappa}~=~ \frac{1}{2 \pi^2} \cdot \int_0^{\infty} q^2 dq
 \Delta V(q) \cdot \int_0^{\infty} dr j_0(qr) [F^2_{n \kappa}+G^2_{n \kappa}]
\label{eq:pert1}
\end{equation}
where $F_{n \kappa}$ and $G_{n \kappa}$ are the small and large components
of the wave function, $n$ is the principle quantum number and $\kappa$
is equal to $-(\ell+1)$ if $j=\ell+\frac{1}{2}$ and $+\ell$ if
$j=\ell-\frac{1}{2}$.    
$\Delta V(q)$ is the Fourier transform of the physical potential.
\begin{equation}
 \Delta V(q)~=~ 4 \pi \cdot \int_0^{\infty} r^2  \cdot  j_0(qr) 
 \cdot \Delta V(r) \,  dr
\label{eq:dv2}
\end{equation}
\begin{equation}
 \Delta V(r)~=~ \frac{1}{2 \pi^2} \cdot \int_0^{\infty} q^2  \cdot  j_0(qr) 
  \cdot \Delta V(q) \, dq
\label{eq:dv1}
\end{equation}
As is well-known \cite{RMP}, 
the Uehling potential in momentum space is given by 
\begin{displaymath}
 V_{Uehl}(q) \,=\, -\frac{4 \alpha(\alpha Z)}{3} \cdot G_E(q) \cdot  
  F(\phi) \,=\, -4 \pi (\alpha Z) \cdot G_E(q) \cdot U_2(q)
\end{displaymath}
\noindent where   $G_E$ is the proton charge form factor, 
$\sinh(\phi) = q/(2 m_e)$ and   
\begin{equation}%
 F(\phi) \,=\, \frac{1}{3} + (\coth^2(\phi)-3) \cdot 
  [1 + \phi \cdot \coth(\phi)]
\label{eq:f-phi}  
\end{equation}%

\noindent $U_2(q)$ is defined in \cite{RMP}.
The vacuum polarization corrections were calculated in momentum space; 
 the formulas (124,125,127) of \cite{RMP}
are completely equivalent to (200) in \cite{eides}.  If the correction
to the transition $2p_{1/2} - 2s_{1/2}$ is calculated in lowest order
perturbation theory using nonrelativistic point Coulomb wave functions,
the result is 205.0074\,meV, in agreement with other
authors\,\cite{eides}. 

The same procedure was used to calculate the two-loop corrections; the
corresponding diagrams were first calculated by K\"allen and Sabry
\cite{kaellen}.  The Fourier transform of the corresponding potential is
given in \cite{RMP,Borie75}.  The result for a point nucleus is
1.5080\,meV. 

 In momentum space 
including the effect of nuclear size on the Uehling potential is
trivial, since the corresponding expression for $\Delta V(q)$ is simply
multiplied by the form factor.  The numbers obtained were the same for a
dipole form factor and for a Gaussian form factor, provided the
parameters were adjusted to reproduce the experimental rms radius of the
proton. The correction can be regarded as taking into account the effect
of finite nuclear size on the virtual electron-positron pair in the
loop.   The contribution of the Uehling potential to the 2p-2s transition
is  reduced by 0.0081\,meV with a proton radius of
0.862\,fm \cite{simon}, and by 0.0085\,meV with a proton radius of
0.880\,fm \cite{rosenfelder}.  
This result is consistent with the number given in \cite{eides}  
(eq.(266)).  More recent values for the proton radius have been given by
Sick \cite{sick} (0.895\,$\pm$\,0.018\,fm) and in the newest CODATA
compilation \cite{codat02} (0.875\,$\pm$\,0.007\,fm). 

\newpage
The numerical values given below were calculated as the
expectation value of the Uehling potential using point-Coulomb Dirac
wave functions with reduced mass: \\
\begin{center} 
    \begin{tabular}{|l|cc|cc|}
   \hline
 &                  point nucleus  &        &    $R_p$=0.875fm  &  \\
 \hline
 & $2p_{1/2}-2s_{1/2}$  & $2p_{3/2}-2s_{1/2}$  & $2p_{1/2}-2s_{1/2}$  &
 $2p_{3/2}-2s_{1/2} $           \\
Uehling          &  205.0282~  & 205.0332~ &   205.0199~ &  205.0250~  \\
Kaellen-Sabry   &    1.50814  &   1.50818  &    1.50807  &  1.50811   \\
   \hline
\end{tabular}        
\end{center}   

The effect of finite proton size calculated here can be  parametrized as
-0.0109$\langle r^2 \rangle$. 
However higher iterations can change these results.  For a very crude
estimate, one can scale previous results for helium \cite{Borie78} and 
assume that the ratio of nonperturbative to perturbative contributions
was the same, giving a contribution of 0.175\,meV.    \\
The contribution due to two and three iterations have been calculated by
\cite{Pachuki1} and \cite{kinoshita}, respectively, giving a total of
0.151\,meV.  An additional higher iteration including finite size and
vacuum polarization is given in ref.\,\cite{Pachuki1} (equations(66) and
(67))  and ref.\,\cite{eides} (equations(264) and (268)).  These amount
to  -0.0164$\langle r^2 \rangle$.    
The best way to calculate this would be an accurate
numerical solution of the Dirac equation in the combined Coulomb-plus
Uehling potential.    

The mixed muon-electron vacuum polarization correction was recalculated
and gave the same result as obtained previously, namely 0.00007\,meV.
\cite{HelvPA,eides}.

The Wichmann-Kroll \cite{WK}  contribution was calculated using the
parametrization for the potential given in \cite{RMP}.  The result
obtained (-0.00103\,meV) is consistent with that given in
\cite{eides}, but not with that given in \cite{Pachuki1}.

The equivalent potential for the virtual Delbr\"uck effect was
recomputed from the Fourier transform given in  \cite{Borie76} and
\cite{RMP}.  The resulting potential was checked by reproducing
previously calculated results for the 2s-2p transition in muonic helium,
and the 3d-2p transitions in muonic Mg and Si.  The result for hydrogen
is +(0.00135 $\pm$ 0.00015)\,meV.  As in the case of muonic helium, this
contribution very nearly cancels the Wichmann-Kroll contribution.  
The contribution  corresponding to three photons to 
the muon and one to the proton  should be analogous to the light by light 
contribution to the muon anomalous moment; to my knowledge, the
corresponding contribution to the muon form factor has never been
calculated.   It will be comparable to the other light by light 
contributions.  For an estimate, the correction to the Lamb shift due
to the contribution to the anomalous magnetic moment was calculated; it
amounts to (-)0.00002\,meV;  the contribution to the muon form factor is
one of the most significant unknown corrections. 

The sixth order vacuum polarization corrections to the Lamb shift in
muonic hydrogen have been calculated by Kinoshita and Nio
\cite{kinoshita}.  Their result for the 2p-2s transition is 
\begin{displaymath}
\Delta E^{(6)} ~=~ 0.120045 \cdot (\alpha Z)^2 \cdot m_r
\left(\frac{\alpha}{\pi}\right)^3 \,\approx \,0.00761 \, \textrm{meV} 
\end{displaymath}
It is entirely possible that the as-yet uncalculated light by light
contribution will give a comparable contribution.

The hadronic vacuum polarization contribution has been estimated by a
number of authors \cite{hadron,friar99,eides}.  It amounts to about
0.012\,meV. 
One point that should not be forgotten about the hadronic VP correction
is the fact that the sum rule or dispersion relation that everyone
(including myself) used does not take into account the fact that the
proton  (nucleus) can 
in principle interact strongly with the hadrons in the virtual hadron
loop. This is irrelevant for the anomalous magnetic moment but probably not 
for muonic atoms.  An estimation of this effect appears to be extremely
difficult, and could easily change the correction by up to 50\%.  
Eides et al. \cite{eides} point out that the graph related to hadronic 
vacuum polarization can also contriibute to the measured value of the
nuclear charge distribution (and polarizability).  It is not easy to
determine where the contribution should be assigned.  

\subsubsection*{Finite nuclear size and nuclear polarization }
\vspace{-0.2cm}
The main contribution due to finite nuclear size has been given
analytically to order ($\alpha Z)^6$ by Friar \cite{friar79}.  The main
result is \\
\begin{equation}
\Delta E_{ns}\,=\,-\frac{2 \alpha Z}{3} \left(\frac{\alpha Z m_r}{n}\right)^3
 \cdot \left[\langle r^2 \rangle - \frac{\alpha Z m_r}{2} \langle r^3
 \rangle_{(2)} +(\alpha Z)^2 (F_{REL}+m^2_r F_{NR})  \right]
\label{eq:FS-friar}
\end{equation}
\noindent where $\langle r^2 \rangle$ is the mean square radius of the 
proton.  
For muonic hydrogen, the coefficient of $\langle r^2 \rangle$ is
5.1975\,(meV fm$^{-2}$), giving an energy shift (for the leading term) 
of 3.862$\pm$0.108\,meV if the proton rms radius is 
0.862$\pm$0.012)\,fm.  
The shift is  4.163$\pm$0.188\,meV
if the proton rms radius is 0.895$\pm$0.018)\,fm, and  3.979$\pm$0.076\,meV
if the proton rms radius of 0.875$\pm$0.007)\,fm. . 
The second term in  Eq.(\ref{eq:FS-friar}) contributes 
-0.0232\,meV for a dipole form factor and -0.0212\,meV for a Gaussian form
factor.  The parameters were fitted to the proton rms radius.  This can
be written as -0.0347$\langle r^2 \rangle^{3/2}$ or  
0.0317$\langle r^2 \rangle^{3/2}$, respectively.  This differs slightly 
from the value given by  Pachucki \cite{Pachuki2}.  The model dependence
introduces an uncertainty of about $\pm$0.002\,meV.
The remaining terms contribute 0.00046\,meV.   This  
estimate includes all of the terms given in \cite{friar79}, while other
authors \cite{Pachuki2} give only some of them.  Clearly the neglected
terms are not negligible. 
There is also a contribution of -$3\cdot10^{-6}$\,meV to the binding
energy of the $2p_{1/2}$-level, and a recoil correction of 
0.012\,meV to the binding energy of the 2s-level.   

As mentioned previously, the finite-size contributions to vacuum
polarization  can be parametrized as  \mbox{$-\,0.0109 \langle r^2 \rangle
\,-\, 0.0164 \langle r^2 \rangle$,}  giving a total of  $-0.0273\langle r^2
\rangle$ or -0.0209(6)\,meV if the proton radius is 0.875\,fm.  

The contribution due to nuclear polarization has been calculated by 
Rosenfelder \cite{rosenfelder99} to be 0.017\,$\pm$\,0.004\,meV, and by
Pachuki \cite{Pachuki2} to be 0.012\,$\pm$\,0.002\,meV.  Other
calculations \cite{srartsev,faustov} give intermediate values
(0.013\,meV and 0.016\,meV, respectively).  The value appearing in 
 table \ref{tab:final} is an average of the three most recent values,
 with the largest quoted uncertainty, which is probably underestimated. 

\subsubsection*{Relativistic Recoil }
\vspace{-0.2cm}
As is well-known, the center-of-mass motion can be separated exactly
from the relative motion only in the nonrelativistic limit.
Relativistic corrections have been studied by many authors, and will not
be reviewed here.  The relativistic recoil corrections summarized in
\cite{RMP} include the effect of finite nuclear size to leading order in
$m_{\mu}/m_N$ properly.

Up to now this method has  been used to treat recoil corrections to
vacuum polarization only in the context of extensive numerical
calculations that include the Uehling potential in the complete
potential,  as described in \cite{RMP}.   They can be included
explicitly, as a perturbation correction to point-Coulomb values.  
Recall that (to leading order in $1/m_N$), the energy levels are given by 
\begin{equation}
  E~=~E_r - \frac{B_0^2}{2  m_N} + \frac{1}{2 m_N} \langle h(r) +  
  2 B_0 P_1(r) \rangle
\label{eq:rmp1}
\end{equation}
where $E_r$ is the energy level calculated using the reduced mass and 
$B_0$ is the unperturbed binding energy.  Also  
\begin{equation}
   h(r)\, = \, - P_1(r)[P_1(r) + \frac{1}{r} Q_2(r)]  
            - \frac{1}{3 r} Q_2(r) [P_1(r) + Q_4(r)/r^3]
\label{eq:rmp2}
\end{equation}
Here 
\begin{align} \label{eq:rmp3}
 P_1(r)&\,=\, 4 \pi \alpha Z \int_r^{\infty} r' \rho(r') dr' 
     & \,=\,&  -V(r)-rV'(r)  \\ \nonumber
Q_2(r)&\,=\,4 \pi \alpha Z \int_0^r r'^2 \rho(r') dr' & \,=\,&  r^2V'(r) \\ \nonumber   
Q_4(r)&\,=\,4 \pi \alpha Z \int_0^r r'^4 \rho(r') dr'   &     &    
\end{align}

An effective charge density $\rho_{VP}$ for vacuum
polarization  can be derived from the Fourier transform of the Uehling
potential. Recall that (for a point nucleus)
\begin{equation*} \begin{split}
V_{Uehl}(r) & \,=\,-\frac{\alpha Z}{r} \frac{2 \alpha}{3\pi} \cdot 
\chi_1(2 m_e r) \\ & \,=\, -(\alpha Z) \frac{2 \alpha}{3\pi} \cdot 
\int_1^{\infty} dz \frac{(z^2-1)^{1/2}}{z^2} \cdot 
\left(1+\frac{1}{2 z^2}\right) \left( \frac{2}{\pi} 
  \int_0^{\infty} \frac{q^2  \cdot  j_0(qr)}{q^2 + 4 m_e^2 z^2}\,dq\right) 
\end{split} \end{equation*}
\noindent where $\chi_n(x)$ is defined in \cite{RMP}.
In momentum space, the Fourier transform of 
$\nabla^2 V$ is obtained by multiplying the Fourier transform of $V$ by
$-q^2$.  
  Note that using the normalizations of \cite{RMP,hyperfine},
one has $\nabla^2 V = - 4 \pi \alpha Z \rho$ where $\rho$ is the 
charge density. 
One then obtains 
\begin{equation} \begin{split}
4 \pi \rho_{VP}(r) & \,=\, \frac{2 \alpha}{3\pi} \cdot 
\int_1^{\infty} dz \frac{(z^2-1)^{1/2}}{z^2} \cdot 
\left(1+\frac{1}{2 z^2}\right) \left(\frac{2}{\pi} 
\cdot \int_0^{\infty} \frac{q^4 \cdot j_0(qr)}{q^2 +4 m_e^2 z^2}\,dq \right) \\
& \,=\, \frac{2}{\pi} \cdot \int_0^{\infty} q^2 U_2(q)  j_0(qr)  \,dq
\label{eq:rhovp}
\end{split} \end{equation}
\noindent  $U_2(q)$ is defined in \cite{RMP}.   
It is also easy to show that 
\begin{equation*} \begin{split}
\frac{d V_{Uehl}}{dr} & \,=\,+\frac{\alpha Z}{r} \frac{2 \alpha}{3\pi} \cdot 
 \left[ \frac{1}{r} \chi_1(2 m_e r) + 2 m_e \chi_0(2 m_e r) \right]  
\\ & \,=\, -\frac{1}{r} V_{Uehl}(r) +(\alpha Z) \frac{2 \alpha}{3\pi} \cdot 
 \frac{2 m_e}{r} \chi_0(2 m_e r)
\end{split} \end{equation*}

Keeping only the Coulomb and Uehling potentials, one finds 
\begin{align*}
 P_1(r) &\,=\,- \alpha Z \frac{2 \alpha}{3\pi} (2 m_e) \chi_0(2 m_e r) \\
 Q_2(r) & \,=\, \alpha Z \left(1 + \frac{2 \alpha}{3\pi}[\chi_1(2 m_e r)
 +  (2 m_e r) \chi_0(2 m_e r)] \right)  \\
 Q_4(r) &\,=\,  \alpha Z \frac{2 \alpha}{3\pi} 
\int_1^{\infty} dz \frac{(z^2-1)^{1/2}}{z^2}  
\left(1+\frac{1}{2 z^2}\right) \\   & \cdot \left( \frac{2}{\pi} \right) 
 \int_0^{\infty} \frac{1}{q^2 + 4 m_e^2 z^2} \frac{[6qr-(qr)^3]\cos(qr)
   + [3(qr)^2-6]\sin(qr)}{q}  \,dq     
\end{align*}
\noindent where $\chi_n(x)$ is defined in \cite{RMP}.  Corrections due
 to finite nuclear size can be  included when a model for the charge
 distribution is given.  This done by Friar \cite{friar79} (and
 confirmed independently for two different model charge distributions); the
 contribution due to finite nuclear size to the recoil correction for
 the binding energy of the 2s-level  is -0.013\,meV.  The factor $1/m_n$
 is replaced by  $1/(m_{\mu}+m_N)$, also consistent with the
 calculations presented in \cite{friar79}.  

Since vacuum polarization is assumed to be a relatively small correction
to the Coulomb potential, it will be sufficient to approximate
$Q_2(r)$ by $\alpha Z/r $.   
After some algebra, one can reduce the expectation values to single
integrals: 
\begin{equation} \begin{split}
\langle  P_1(r) \rangle \,=\, & 2 m_e \alpha Z \frac{2 \alpha}{3 \pi}
 \int_1^{\infty} \frac{(z^2-1)^{1/2}}{z}\cdot \left(1+\frac{1}{2
 z^2}\right) \cdot \\ & ~~~~~~~~~~~~~~~~~\left(\frac{(az)^2-az+1}{(1+az)^5}  
 \delta_{\ell 0} +\frac{1}{(1+az)^5}\delta_{\ell 1} \right)
 \,dz 
\label{eq:p1}
\end{split} \end{equation}

\begin{equation} \begin{split}
\Big\langle \frac{\alpha Z}{r} P_1(r) \Big\rangle \,=\,
 &   - (\alpha Z)^3  m_r m_e \frac{2 \alpha}{3 \pi}
 \int_1^{\infty} \frac{(z^2-1)^{1/2}}{z}\cdot \left(1+\frac{1}{2
 z^2}\right) \cdot \\ & ~~~~~~~~~~~~~~\left(\frac{2(az)^2+1}{2 (1+az)^4}  
 \delta_{\ell 0} +\frac{1}{2(1+az)^4}\delta_{\ell 1} \right)
 \,dz 
\label{eq:azp1}
\end{split} \end{equation}
\noindent  with $a \,=\,2 m_e /(\alpha Z m_r)$.  
When Eq.\,(\ref{eq:p1}) is multiplied by $-2 B_0/(m_{\mu}+m_N)$ this
results in a shift of $-0.00015$\,meV for the 2s-state and of
$-0.00001$\,meV for the 2p-state, and 
when Eq.\,(\ref{eq:azp1}) is multiplied by $1/(m_{\mu}+m_N)$ this
results in a shift of 
$0.00489$\,meV for the 2s-state and of $0.00017$\,meV for the 2p-state.
These expectation values also appear when vacuum polarization is
included in the Breit equation \cite{Pachuki3}.

Finally,
\begin{equation} \begin{split}
\Big\langle \frac{\alpha Z}{3 r^4} Q_4(r) \Big\rangle \,=\,
 &   - \frac{(\alpha Z)^4  m_r^2}{6} \frac{2 \alpha}{3 \pi}
 \int_1^{\infty} \frac{(z^2-1)^{1/2}}{z^2}\cdot \left(1+\frac{1}{2
 z^2}\right) \cdot \\ & ~~~~~~~~~~~~~\Biggl[ \Bigl[-\frac{6}{az}
 \Bigl(\frac{2 +az}{1+az}-\frac{2}{az} \ln(1+az) \Bigr) +
 \frac{3(az)^2+2az-1}{(1+az)^3} + \\ &~~~~~~~~~~~~ \frac{3+az}{4(1+az)^4} \Bigr]  
 \delta_{\ell 0} + \frac{1-3az-2(az)^2}{4(1+az)^4} \delta_{\ell 1} \Biggr]
 \,dz 
\end{split} \end{equation}
When multiplied by $1/(m_{\mu}+m_N)$ this results in a shift of
0.002475\,meV for the 2s-state and of 0.000238\,meV for the 2p-state.

Combining these expectation values according to equations \ref{eq:rmp1}
and \ref{eq:rmp2}, one finds a contribution to the 2p-2s transition of 
-0.00419\,meV.  To obtain the full relativistic and recoil corrections,
one must add the difference between the expectation values of the
Uehling potential calculated with relativistic and nonrelativistic wave
functions, 
giving a total correction of 0.0166\,meV.  This is in fairly 
good agreement with the correction of .0169\,meV calculated by Veitia
and Pachucki \cite{Pachuki3}, using a generalization of the Breit
equation \cite{barker} which is similar to that given in \cite{hyperfine}.  
The treatment presented here has the
advantage of avoiding second order perturbation theory.

The review by Eides et.al \cite{eides} gives a better version of the two
photon recoil (Eq. 136) than was available for the review by Borie and
G. Rinker \cite{RMP}.  Evaluating this expression for muonic hydrogen
gives a contribution of -0.04497\,meV to the 2p-2s transition.
Higher order radiative recoil corrections give an additional
contribution of -0.0096\,meV \cite{eides}. 
However, some of the contributions to the expressions given in
\cite{eides} involve logarithms of the mass ratio $m_{\mu}/m_N$.
Logarithms can only arise in integrations in the region from 
$m_{\mu}$ to $m_N$; in this region the effect of the nuclear form factor
should not be neglected.  Pachucki \cite{Pachuki1} has estimated a
finite size correction to this of about 0.02\,meV, which seems to be
similar to the term proportional to $\langle r^3 \rangle_{(2)} $ given
in Eq.(\ref{eq:FS-friar}) as calculated in the external field
approximation by Friar \cite{friar79}.  This two-photon correction
requires further investigation.  In particular, the parametrization of
the form factors used in any calculation should reproduce the correct
proton radius.    

An additional recoil correction for states with $\ell \, \ne \, 0$ has
been given by \cite{barker} (see also \cite{eides}).  It is 
\begin{equation}
\Delta E_{n, \ell ,j}~=~\frac{(\alpha Z)^4 \cdot m_r^3}{2 n^3 m_N^2} 
 (1-\delta_{\ell 0}) \left(\frac{1}{\kappa (2\ell+1)}\right)
\label{eq:recoil1}
\end{equation}
When evaluated for the 2p-states of muonic hydrogen, one finds a
contribution to the 2p-2s transition energy of 0.0575\,meV for the
2p$_{1/2}$ state and -0.0287\,meV for the 2p$_{3/2}$ state.

\subsubsection*{Muon Lamb Shift }
\vspace{-0.2cm}
For the calculation of muon self-energy and vacuum polarization, the
lowest order (one-loop approximation) contribution is well-known, at
least in perturbation theory.
Including also  muon vacuum polarization
(0.0168\,meV) and an extra term of order $(Z \alpha)^5$ as given in
\cite{eides}:
\begin{displaymath}
\Delta E_{2s}\,=\, \frac{\alpha (\alpha Z)^5 m_{\mu} }{4} \cdot
\left(\frac{m_r}{m_{\mu}}\right)^3   \cdot    
 \left(\frac{139}{64} + \frac{5}{96} - \ln(2) \right)
\end{displaymath}
which contributes -0.00443\,meV, one finds a  contribution of
-0.66788\,meV for the \mbox{$2s_{1/2}-2p_{1/2}$} transition and 
 -0.65031\,meV for the  \mbox{$2s_{1/2}-2p_{3/2}$} transition.

A misprint in the evaluation of the contribution of the higher order muon 
form factors (contributing to the fourth order terms) has been
corrected.  The extra electron loop contribution 
to $F_2$(0) is should be 1.09426$(\alpha/\pi)^2$.  This reproduces 
the correct coefficient of $(\alpha/\pi)^2$ from the muon (g-2)
analyses. This is .7658, which is equal to 1.09426-0.32848. 

The fourth order electron loops \cite{barbieri73} dominate the  
 fourth order contribution (-0.00169\,meV and  -0.00164\,meV,
respectively).   The rest is the same as for the electron \cite{RMP}.  
The contribution of the electron loops alone is 
 -0.00168\,meV for the $2s_{1/2}-2p_{1/2}$ transition and  -0.00159\,meV
for the $2s_{1/2}-2p_{3/2}$ transition. 

Pachuki \cite{Pachuki1} has estimated an additional contribution of
-0.005\,meV for a contribution corresponding to a vacuum polarization
insert in the external photon. 

\subsubsection*{Summary of contributions }
\vspace{-0.2cm}
Using the fundamental constants from the CODATA 1998 (\cite{Mohr})  
one finds the  transition energies in meV in table \ref{tab:final}.  Here
the main  vacuum polarization contributions are given for a   
point nucleus, using the Dirac equation with reduced mass. 
  Some uncertainties have been increased from the values
given by the authors, as discussed in the text. 

The finite size corrections up to order $(\alpha Z)^5$ can be
parametrized as \\
 \mbox{$ 5.1975 \langle r^2 \rangle \,-\,0.0109 \langle
 r^2 \rangle\,-\, 0.0164 \langle r^2 \rangle\,+\,0.0347 \langle r^3
 \rangle_{(2)}$.}  
\begin{table}[!h]
\begin{center} 
    \begin{tabular}{|lrr|}
      \hline
Contribution  &  Value (meV) &  Uncertainty (meV) \\
      \hline
Uehling        &     205.0282~    &  \\
K\"allen-Sabry     &       1.5081~  &           \\
Wichmann-Kroll   &      -0.00103     & \\
virt. Delbrueck    &     0.00135   &  0.00015\\
mixed mu-e VP    &       0.00007   &  \\
hadronic VP       &      0.011~~~    &   0.002~~~     \\       
sixth order \cite{kinoshita}  & 0.00761    &   \\
\hline
recoil  \cite{eides} (eq136)     & -0.04497  &  \\
recoil, higher order \cite{eides} & -0.0096~ & \\
recoil, finite size \cite{friar79} & 0.013~~~  &  0.001~~~ \\
recoil correction  to VP \cite{RMP}  &  -0.0041~~ &  \\
additional recoil  \cite{barker} & 0.0575~~ &   \\
\hline
muon Lamb shift   &    &  \\
second order      &     -0.66788 & \\
fourth order      &     -0.00169 &  \\
\hline
nuclear size  ($R_p$=0.875\,fm)  & &  0.007\,fm \\
main correction \cite{friar79}  & -3.979~~    &  0.076~~~   \\
order $(\alpha Z)^5$ \cite{friar79} &  0.0232~  &  0.002~~~ \\
order $(\alpha Z)^6$ \cite{friar79} &  -0.0005~  &  \\
correction to VP  &   -0.0083~  &  \\
polarization      &    0.015~~~ &  0.004~~          \\
\hline
Other (not checked)  &    &  \\
VP iterations \cite{Pachuki1}  &  0.151~~~  &    \\
VP insertion in self energy \cite{Pachuki1}  &  -0.005~~~  &    \\ 
additional size for VP   \cite{eides}  &  -0.0128~~  &    \\ 
\hline
\end{tabular}        
 \caption{ Contributions to the muonic hydrogen Lamb shift. The proton
   radius is taken from \cite{codat02}. }
  \label{tab:final}
\end{center}   
\end{table}        
\begin{center}
 \begin{minipage}{\linewidth}
In the case of the muon Lamb shift, the numbers in table\,\ref{tab:final}
are for the $2s_{1/2}-2p_{1/2}$ transition.   The corresponding numbers
for the  
$2s_{1/2}-2p_{3/2}$ transition are -0.65031\,meV and -0.00164\,meV,
respectively. 
\end{minipage}
\end{center}

\subsubsection*{Fine structure of the 2p state }
\vspace{-0.2cm}
There are two possible ways to calculate the fine structure.  One is to
start with the point Dirac value, include the contribution due to vacuum
polarization, as calculated above, as well as the spin-orbit splitting
(computed perturbatively) due to the muon's anomalous magnetic moment, 
and recoil as given by Eq.(\ref{eq:recoil1}).  The results are
summarized in table \ref{tab:FS}. 
\begin{table}[!h]
\begin{center} 
    \begin{tabular}{|lr|}
      \hline
          &  \mbox{$ E(2p_{3/2}) - E(2p_{1/2})$ (meV)} \\
Dirac          &      8.41564                      \\
Uehling(VP)        &     0.0050~      \\
K\"allen-Sabry      &     0.00004  \\
\hline
anomalous moment $a_{\mu}  $ &          \\
second order      &     0.01757  \\
higher orders      &     0.00007  \\
\hline
Recoil  (Eq.(\ref{eq:recoil1}))     &  -0.0862~    \\
\hline
Total Fine Structure    &                   8.352~~   \\
\hline
\end{tabular}        
 \caption{ Contributions to the fine structure of the 2p-state in muonic
   hydrogen.  }
  \label{tab:FS}
\end{center}   
\end{table}

An alternative method is to use the formalism given in \cite{hyperfine}
(and elsewhere, see, eg. \cite{barker,eides}) which gives the energy 
shift as the expectation value of 
\begin{equation}
- \frac{1}{r}\,\frac{dV}{dr} \cdot \frac{1+a_{\mu}+(a_{\mu}+1/2)m_N/m_{\mu}}
{m_N m_{\mu}} \vec{L} \cdot \vec{\sigma}_{\mu}
\label{eq:fs1}
\end{equation}%
Note that 
\begin{displaymath}
 \frac{1}{m_N m_{\mu}}+\frac{1}{2 m_{\mu}^2}\,=\,
  \frac{1}{2 m_r^2}-\frac{1}{2 m_N^2}
\end{displaymath}
so that the terms not involving  $a_{\mu}$ in the spin-orbit
contribution are really the Dirac fine
structure plus the Barker-Glover correction (Eq. \ref{eq:recoil1}) 

The Uehling potential has to be included in the potential $V(r)$.  
For states with $\ell\,>\,0$ in light atoms, and
neglecting the effect of finite nuclear size, we may take 
\begin{equation}
\frac{1}{r} \frac{d V}{dr}~=~\frac{ \alpha Z}{r^3} \cdot \left[ 1 +
 \frac{2 \alpha}{3 \pi} \int_1^{\infty} \frac{(z^2-1)^{1/2}}{z^2}\cdot
\left(1+\frac{1}{2 z^2}\right) \cdot (1 + 2 m_e r z) \cdot e^{-2 m_e r z} \,dz \right]
\label{eq:vp2p}
\end{equation}
which is obtained from the Uehling potential \cite{Uehling,serber}
by differentiation.  Then, assuming that it is sufficient to use
nonrelativistic point Coulomb wave functions for the 2p state, one finds 
\begin{displaymath}
\Big\langle \frac{1}{r^3} \Big\rangle_{2p} \rightarrow 
 \Big\langle \frac{1}{r^3} \Big\rangle_{2p} \cdot (1+\varepsilon_{2p})
\end{displaymath}
\noindent  where 
\begin{equation}
\varepsilon_{2p} ~=~ \frac{2 \alpha}{3 \pi} 
 \int_1^{\infty} \frac{(z^2-1)^{1/2}}{z^2}\cdot \left(1+\frac{1}{2 z^2}\right)
  \cdot \left(\frac{1}{(1+az)^2} + \frac{2 az}{(1+az)^3}\right) \,dz
\label{eq:eps2p}
\end{equation}

\noindent  with $a \,=\,2 m_e /(\alpha Z m_r)$.  
The result for the fine structure is 
\begin{equation}
\frac{-(\alpha Z)^4 m_r^3}{ n^3 (2\ell +1) \kappa}\cdot 
\left(\frac{1}{m_N m_{\mu}}+\frac{1}{2 m_{\mu}^2}+\frac{a_{\mu}}{m_{\mu} m_r}\right)  
 \cdot (1 + \varepsilon_{2p}) 
\label{eq:fs2}
\end{equation} 
\noindent where $\varepsilon_{2p}$ is given by  Eq.(\ref{eq:eps2p}).
In this case, the terms involving $a_{\mu}$ in the expression for the
muon Lamb shift are included, and should not be double counted.   With a
numerical value of $\varepsilon_{2p}$\,=\,0.000365, one finds a
contribution of 0.00305\,meV (compared with 0.005\,meV using Dirac wave
functions).  

Numerically, the terms not involving $a_{\mu}$ give a contribution of
8.3291\,meV and the contribution from $a_{\mu}$ gives a contribution of
0.0176\,meV, for a total of 8.3467\,meV, in good agreement with Eq.\,80 of 
\cite{Pachuki1}.   When the vacuum polarization correction is added, the
result is  only very slightly  
different from the Dirac value of 8.352\,meV.
The contribution due to the  anomalous magnetic moment of the muon is
the same in both cases. 

In both cases one should include the 
$B^2/2M_N$-type correction to the fine structure. (see \cite{eides},
Eq(38)).  This is tiny 
($5.7 \cdot 10^{-6}$\,meV) and is not included in the table.  Friar
\cite{friar79} has given  expressions for the energy shifts of the
2p-states due to finite nuclear size.  These were calculated and found
to give a negligible contribution ($3.1 \cdot 10^{-6}$\,meV) to the fine
structure of the 2p-state. 

\subsubsection*{Hyperfine structure }
\vspace{-0.2cm}
The hyperfine structure is calculated in the same way  as was done in
earlier work \cite{hyperfine,BorieHe3}, but with improved accuracy.  
Most of the formalism and results are similar to those given by
\cite{Pachuki1}. \\

\noindent {\bf The 2p state: } \\ 
The hyperfine structure of the 2p-state is given by \cite{hyperfine}
($F$ is the total angular momentum of the state)
\begin{equation} \begin{split}
\frac{1}{4m_{\mu} m_N} \Big\langle \frac{1}{r}\,\frac{dV}{dr}\Big\rangle_{2p}
 & \cdot(1+\kappa_p) \biggl[2(1+x) \delta_{j j'} (F(F+1)-11/4) \\
 & + 6 \hat{j} \hat{j}' (C_{F1}(1+a_{\mu})-2(1+x)) \left\{ \begin{array}{ccc} 
 \ell & F & 1 \\
 \frac{1}{2} & \frac{1}{2} & j 
 \end{array} \right\}  \left\{  \begin{array}{ccc} 
 \ell & F & 1 \\
 \frac{1}{2} & \frac{1}{2} & j' 
 \end{array} \right\}     \biggr] 
\end{split}
\end{equation}
\noindent where $\hat{j} = \sqrt{2 j + 1}$, the 6-j symbols are defined
 in \cite{edmonds},  and \\
 $C_{F1}=\delta_{F1}-2\delta_{F0}-(1/5)\delta_{F2}$. 



\begin{displaymath}
 x\,=\, \frac{m_{\mu} (1 + 2 \kappa_p)}{2 m_N (1 + \kappa_p)}
\end{displaymath}
\noindent  represents a recoil correction due to Thomas precession
\cite{hyperfine,barker}.
The same correction due to vacuum polarization (Eq.\,(\ref{eq:eps2p})) 
should be applied to the HFS shifts of the 2p-states, as well as to the 
spin-orbit term.    

As has been known for a long time \cite{hyperfine,BorieHe3,Pachuki1},
the states with total angular momentum $F=1$ are a superposition of the
states with $j=1/2$ and $j=3/2$.  
Let the fine structure splitting be denoted by 
$\delta \,=\, E_{2p3/2} - E_{2p1/2} $, and let 
\begin{displaymath}
\beta\,=\,\frac{(\alpha Z)^4 m_r^3}{3m_{\mu} m_N}
 \cdot (1 + \kappa_p) 
\end{displaymath}
\noindent  and $\beta ' \,=\, \beta \cdot (1+ \varepsilon_{2p})$.

The energy shifts of the 2p-states with total angular momentum F
(notation $^{2F+1}L_j$) are then given in table \ref{tab:HFS-2p} 
\begin{table}[!h]
\begin{center} 
    \begin{tabular}{lrr}
State  &          Energy         &   Energy in meV  \\
     \hline
$^1 p_{1/2}$    & -$\beta ' (2+x + a_{\mu})/8 $~    &   -5.971       \\
$^3 p_{1/2}$    & $(\Delta - R)/2$~    &     1.846        \\
$^3 p_{3/2}$    & $(\Delta + R)/2$~     &    6.376         \\
$^5 p_{3/2}$    & $\delta + \beta ' (1+5x/4 - a_{\mu}/4)/20$   & 9.624   \\
\end{tabular}        
 \caption{ Hyperfine  structure of the 2p-state in muonic
   hydrogen.  }
  \label{tab:HFS-2p}
\end{center}   
\end{table}    
\begin{center}     
 \begin{minipage}{\linewidth}
\noindent   where
\begin{displaymath}
\Delta\,=\, \delta - \beta '(x - a_{\mu}) /16
\end{displaymath}
\begin{displaymath}
 R^2 \,=\, [\delta  - \beta '(1 +7x/8 + a_{\mu}/8) /6]^2
 + (\beta')^2 (1+2x-a_{\mu})^2/288  
\end{displaymath}
\end{minipage}
\end{center}   
(Here $\delta$ = 8.352\,meV) 
Some minor errors in \cite{hyperfine} have been corrected.  These
numbers differ slightly from those given in ref.\,\cite{eides}. \\

\noindent {\bf The 2s-state: } \\ 
The basic hyperfine splitting of the 2s-state is given by 
\begin{displaymath}
\Delta \nu_F\,=\,\frac{(\alpha Z)^4 m_r^3}{3m_{\mu} m_N}
 \cdot (1 + \kappa_p) \cdot (1 + a_{\mu}) \,=\,\beta \cdot (1 + a_{\mu}) 
 \,=\,22.8332\,meV
\end{displaymath}
(see, for example \cite{eides}  , Eq. (271,277))
As was shown in \cite{hyperfine,eides}, the energy shift of the 2s-state 
is given by: 
\begin{equation}
\Delta E_{2s}~=~\beta \cdot (1 + a_{\mu}) \cdot (1+\varepsilon_{VP} +
 \varepsilon_{vertex} + \varepsilon_{Breit} + \varepsilon_{FS,rec}) 
 \cdot [\delta_{F1}-3\delta_{F0}]/4  
\label{eq:hf2s}
\end{equation}

\noindent  Here (\cite{Brodsky})
\begin{displaymath}
\varepsilon_{vertex} \,=\, \frac{2 \alpha (\alpha Z)}{3} 
\left(\ln(2)-\frac{13}{4}\right) \,=\, -1.36 \cdot 10^{-4}
\end{displaymath}
\noindent  and (\cite{eides}, Eq.\,(277))
\begin{displaymath}
\varepsilon_{Breit} \,=\, \frac{17 (\alpha Z)^2}{8} 
  \,=\, 1.13 \cdot 10^{-4}
\end{displaymath}
The vacuum polarization correction has two contributions.  One of these
is a result of a modification of the magnetic interaction between the
muon and the nucleus and is given by (see \cite{BorieHe3})
\begin{equation} \begin{split}
\varepsilon_{VP1} \,=\,& \frac{4 \alpha}{3 \pi^2}  
 \int_0^{\infty} r^2 \,dr \left(\frac{R_{ns}(r)}{R_{ns}(0)}\right)^2
 \int_0^{\infty} q^4 j_0(qr) G_M(q) \,dq   \\
 & \int_1^{\infty} \frac{(z^2-1)^{1/2}}{z^2}\cdot
\left(1+\frac{1}{2 z^2}\right) \cdot \frac{dz}{4 m_e^2 [z^2 + (q/2 m_e)^2]} 
\end{split}
\end{equation}
One can do two of the integrals analytically and obtains for the
2s-state (with
\mbox{$a=2m_e/(\alpha Z m_r)$} and  \mbox{$\sinh(\phi) = q/(2 m_e) = K/a$})  
\begin{equation}
\varepsilon_{VP1} \,=\, \frac{4 \alpha}{3 \pi^2}  
 \int_0^{\infty} \frac{K^2}{(1+K^2)^2} F(\phi) G_M(\alpha Z m_r K) \,dK 
\left[2 -\frac{7}{(1+K^2)}+\frac{6}{(1+K^2)^2}\right]
\label{eq:epsvp1}
\end{equation}
where $F(\phi)$ is known from the Fourier transform of the Uehling
potential and is given by Eq(\ref{eq:f-phi}).

The other contribution, as discussed by \cite{Brodsky,sternheim} arises
from the 
fact that the lower energy hyperfine state, being more tightly bound,
has a higher probability of being in a region where vacuum polarization
is large.  This results in an additional energy shift of  
\begin{displaymath}
 2 \int V_{Uehl}(r) \psi_{2s}(r) \delta_M \psi_{2s}(r) d^3r
\end{displaymath}
Following Ref.\,\cite{Brodsky} with $y=(\alpha Z m_r/2) \cdot r$, one has 
\begin{displaymath}
   \delta_M \psi_{2s}(r) \,=\, 2 m_{\mu} \Delta \nu_F \psi_{2s}(0) 
\left(\frac{2}{\alpha Z m_r}\right)^2 \exp(-y) 
\left[(1-y)(\ln(2y)+\gamma)+\frac{13y-3-2y^2}{4}-\frac{1}{4y}\right]
\end{displaymath}
\noindent ($\gamma$ is Euler's constant), and
\begin{displaymath}
  \psi_{2s}(r)\, =\, \psi_{2s}(0) (1-y) \exp(-y)
\end{displaymath}
 One finds after a lengthy integration
\begin{multline}
\varepsilon_{VP2} \,=\, \frac{16 \alpha}{3 \pi^2} 
 \int_0^{\infty} \frac{dK}{1+K^2}  G_E(\alpha Z m_r K)  F(\phi) \\
\biggl\{
 \frac{1}{2}-\frac{17}{(1+K^2)^2}+\frac{41}{(1+K^2)^3}-\frac{24}{(1+K^2)^4} \\
+\frac{\ln(1+K^2)}{1+K^2} \left[2-\frac{7}{(1+K^2)}+\frac{6}{(1+K^2)^2}\right]
 \\  + \frac{\tan ^{-1}(K)}{K}
 \biggl[1-\frac{19}{2(1+K^2)}+\frac{20}{(1+K^2)^2}-\frac{12}{(1+K^2)^3}\biggr] 
 \biggr\}
\label{eq:epsvp2}
\end{multline}
Sternheim\cite{sternheim} denotes the two contributions by $\delta_M$
and  $\delta_E$, respectively.
An alternative exression, obtained by assuming a point nucleus, using
Eq.(131)  from \cite{RMP} for the Uehling potential, and doing the
integrations in  a different order, is  
\begin{equation} \begin{split}
\varepsilon_{VP2} \,=\,& \frac{16 \alpha}{3 \pi} 
 \int_1^{\infty} \frac{(z^2-1)^{1/2}}{z^2}\cdot
  \left(1+\frac{1}{2 z^2}\right) \cdot \frac{1}{(1+az)^2}   \\
& \times \biggl[\frac{az}{2}-\frac{1}{1+az}+\frac{23}{8(1+az)^2} -
 \frac{3}{2(1+az)^3}    \\ 
&  +\ln(1+az)\cdot\left(1-\frac{2}{1+az}+\frac{3}{2(1+az)^2}\right)\biggr] 
\,dz 
\end{split}
\end{equation}
\noindent  with $a \,=\,2 m_e /(\alpha Z m_{red})$.
Both methods give the same result. \\
In the case of ordinary hydrogen, each of these contributes 
$3 \alpha^2/8 = 1.997 \cdot 10^{-5}$.  The accuracy of the numerical
integration was checked by reproducing these results.
One can thus expect that muonic vacuum polarization will contribute 
$3 \alpha^2/4 \simeq 4 \cdot 10^{-5}$, as in the case of normal
hydrogen.  This amounts to an energy shift of 0.0009\,meV.  
Contributions due to the weak interaction or hadronic vacuum
polarization should be even smaller.  
For muonic hydrogen, one obtains 
 $\varepsilon_{VP1}$=0.00211 and  $\varepsilon_{VP2}$=0.00325 for a
 point nucleus. Including the effect of the proton
 size (with $G_E(q)=G_M(q)$ as a dipole form factor) reduces these
 numbers to  0.00206 and 0.00321, respectively.    For the
 case of muonic $^3He$ \cite{BorieHe3}, the corresponding numbers are 
 $\varepsilon_{VP1}$=0.00286 and  $\varepsilon_{VP2}$=0.00476.
The contribution to the hyperfine splitting of the 2s-state is then  
0.0470\,meV+0.0733\,meV=0.1203\,meV (0.1212\,meV if muonic vacuum
 polarization is included).  The combined Breit and vertex
 corrections reduce this value to 0.1207\,meV.  (0.1226\,meV if the
 proton form factors are not taken into account). 

The contribution to the hyperfine structure from the two loop diagrams     
\cite{kaellen} can be calculated by replacing 
$U_2(\alpha Z m_r K) = (\alpha / 3\pi)  F(\phi) $ by 
$U_4(\alpha Z m_r K)$ (as given in \cite{RMP,Borie75}) in equations 
\ref{eq:epsvp1} and \ref{eq:epsvp2}.  The resulting contributions are  
$1.64 \cdot 10^{-5}$ and $2.46 \cdot 10^{-5}$, respectively, giving a
total shift of 0.0009\,meV.

The correction due to finite size and recoil have been given in
\cite{Pachuki1} as -0.145\,meV, while a value of -0.152\,meV is given in
\cite{martynenko}.   Ref.\,\cite{Pachuki1} also gives 
 a correction as calculated by Zemach
(\cite{zemach}) equal to -0.183\,meV, but claims that this correction
does not treat recoil properly.   The Zemach correction is equal to 
\begin{displaymath}
\varepsilon_{Zem} \,=\, -2  \alpha Z m_r \langle r \rangle_{(2)}
\end{displaymath}
where  $\langle r \rangle_{(2)}$ 
is given in \cite{hyperfine,friar79,friar04}.  Using the value 
$\langle r \rangle_{(2)}\,=\,1.086 \pm 0.012$\,fm from \cite{friar04},
gives   $\varepsilon_{Zem}\,=\,-0.00702$, and a contribution of 
of \mbox{-0.1742\,meV} to the hyperfine splitting of the 2s state.
Including this, but not other recoil corrections to the hyperfine
structure of the 2s-state gives a total splitting of 22.7806\,meV.
Additional higher order corrections calculated in
Ref.\,\cite{martynenko} amount to a total of -0.0003\,meV and are not
included here.  


\subsubsection*{Summary of contributions and Conclusions}
\vspace{-0.2cm}
 The most important contributions to the Lamb shift in muonic hydrogen,
including hyperfine structure, have been independently recalculated. A new
calculation of  some terms that were
omitted in the most recent literature, such as the virtual Delbr\"uck 
effect \cite{Borie76} and an alternative calculation of the relativistic
recoil correction have been  presented. 

Numerically the results given in  table \ref{tab:final} add up to a
total correction of \\
 \mbox{(206.032(6)  - 5.225\,$\langle r^2 \rangle$ 
+ 0.0347\,$\langle r^2 \rangle^{3/2}$)\,meV\,=\,202.055$\pm$0.12\,meV.}    
(for the value of the proton radius from \cite{codat02}).
As is well known, most of the uncertainty arises from the uncertainty in 
the proton radius. 

However, the contribution of the light-by-light graph to the muon form
factor has  not yet been calculated.  Also, since $m_{\mu}/m_p\,=0.1126$
is much larger than $\alpha Z$, it is possible that recoil corrections
of higher order in the mass ratio, that have never been calculated,
could be significant at the level of the expected experimental accuracy
of about $0.01$\,meV.  In particular, the two-photon recoil
corrections, including finite nuclear size, should be recalculated to
resolve (small) inconsistencies among various theoretical results.
\medskip

\subsubsection*{ Acknowledgments }
\vspace{-0.2cm}
The author wishes to thank M. Eides, E.-O. Le~Bigot and F. Kottmann for
extensive email correspondence regarding this work.

\small
\def\refname{{\normalsize References}}

\end{document}